# Orbital Symmetry of Ba(Fe$_{1-x}$Co$_x$)$_2$As$_2$ Superconductors Probed with X-ray Absorption Spectroscopy

C. Parks Cheney,[1] F. Bondino,[2] T. A. Callcott,[1] P. Vilmercati,[1] D. Ederer,[3] E. Magnano,[2] M. Malvestuto,[4] F. Parmigiani,[2,5] A. S. Sefat,[6] M. A. McGuire,[6] R. Jin,[6] B. C. Sales,[6] D. Mandrus,[6] D. J. Singh,[6] J. W. Freeland,[7] and N. Mannella[1*]

[1] University of Tennessee, Department of Physics and Astronomy, Knoxville, TN 37996-1200, USA

[2] IOM CNR Laboratorio TASC, S.S. 14, km 163.5, I-34012 Trieste, Italy

[3] Tulane University, Department of Physics, New Orleans, LA 70118 USA

[4] Sincrotrone Trieste S.C.p.A., Area Science Park, S.S. 14, km 163.5, I-34012 Trieste, Italy

[5] Dipartimento di Fisica, Università degli Studi di Trieste, I-34127, Trieste, Italy

[6] Oak Ridge National Laboratory, Oak Ridge, TN 37831, USA

[7] Advance Photon Source, Argonne National Laboratory, Argonne, IL 60439, USA

*Corresponding author. E-mail address: nmannell@utk.edu.

# ABSTRACT

The orbital symmetries of electron doped iron-arsenide superconductors Ba(Fe$_{1-x}$Co$_x$)$_2$As$_2$ have been measured with x-ray absorption spectroscopy. The data reveal signatures of Fe d electron itinerancy, weak electronic correlations, and a high degree of Fe-As hybridization related to the bonding topology of the Fe d$_{xz+yz}$ states, which are found to contribute substantially at the Fermi level. The energies and detailed orbital character of Fe and As derived unoccupied *s* and *d* states are found to be in remarkably good agreement with the predictions of standard density functional theory.

**INTRODUCTION**

The discovery of superconductivity with critical temperature ($T_C$) exceeding 55 K in the iron-pnictide (FeSC) compounds has offered the community a new set of materials hosting high temperature superconductivity, thus breaking the exclusivity of cuprates [1,2,3,4,5]. The FeSC are multi-orbital systems, with a complex Fermi surface consisting of different bands originating from the hybridization of the Fe $d$ orbitals. The symmetry of the orbitals and their hybridization play a central role in the interplay between magnetism and superconductivity and in the nature of the superconducting pairing [6,7,8,9]. This is the case even in a weak coupling band structure framework, where the Lindhard function that determines the nesting between hole and electron Fermi surfaces contains an overlap matrix element that is strong only when the orbital characters match. This nesting plays a central role in theories relying on interband pairing, and also in the nesting that may underlie the spin-density wave antiferromagnetism. In more strongly coupled theories, Hund's coupling and Coulomb repulsions in general provide inter-orbital interactions that may change the orbital characters from band structure results, modify the pairing interactions and open additional pairing channels [8,9]. The experimental determination of the orbital symmetry in FeSC is thus of fundamental importance. Very recently, two different groups have investigated the orbital character of the Fe $d$ bands in proximity of the Fermi level with angle resolved photoemission (ARPES) and highlighted the complexity of this issue, which still remains a controversial subject [10,11].

Here, we address the symmetry and topology of the unoccupied orbitals with elemental sensitivity by reporting the results of polarization-dependent XAS measurements of the electronic structure in the normal state of $BaFe_2As_2$ and $BaFe_{1.8}Co_{0.2}As_2$ single crystals. Polarization dependent XAS is a premiere tool for measuring bonding topology and symmetry of unoccupied orbitals. The electric field of linearly polarized x-rays acts as a search light for the direction of chemical bonds of the atom selected by its absorption edge [12]. Our analysis indicates that the Fe $d_{xz+yz}$ states contribute substantially to the density of states at the Fermi level and are hybridized with As $p_{x+y}$ states.

**METHODS**

The Ba(Fe$_{1-x}$Co$_x$)$_2$As$_2$ materials crystallize in high quality single crystals with the tetragonal ThCr$_2$Si$_2$ structure-type (space group I4/mmm), and have a maximum $T_C$ = 22 K for x = 0.1 nominal doping [13]. The parent compound BaFe$_2$As$_2$ exhibits a magnetic and structural transition at $T_{ST} \approx$ 140 K, below which the crystal symmetry is reduced (orthorhombic, Fmmm space group) and long-range antiferromagnetic ordering sets in [14]. The *a* and *b* axis of the orthorhombic unit cell are not equivalent, with the Fe moments aligning antiferromagnetically and ferromagnetically along the longer *a* and shorter *b* axes, respectively. No structural phase transition and long range magnetic ordering occur in the optimally doped material [13]. High-quality single crystals were grown out of FeAs flux according to modalities described elsewhere [13]. Electron probe microanalysis carried out with a JEOL JSM-840 scanning electron microprobe on several crystals indicated that, for the x = 0.1 nominal doping, 8.0(5)% of the Fe is replaced by Co in BaFe$_2$As$_2$. This material will be referred to below by the nominal BaFe$_{1.8}$Co$_{0.2}$As$_2$.

Density functional calculations were performed within the local density approximation using the general potential linearized augmented planewave (LAPW) method as implemented in the Wien2k code, without including magnetism. The orbital characters were obtained by projections onto the LAPW spheres of radius 2.1. For both Fe and As, the calculated spectra were obtained including the dipole matrix elements to the unoccupied s and d states without correction for final state (core hole) effects.

XAS measurements at the Fe and As $L_{2,3}$ edges were carried out at the BACH beamline at Elettra (Trieste, Italy), the soft x-ray fluorescence end-station of beamline 8.0.1 of the Advanced Light Source (ALS), and the beamline 4-ID-C at the Advanced Photon Source (APS). Several samples from different batches were measured in total electron yield (TEY) and total fluorescence yield (TFY) both above $T_{ST}$ (200 K) and below $T_{ST}$ (50 K and 77 K). The spectra have been recorded from samples both in pristine conditions and after being cleaved *in-situ* in ultrahigh vacuum (UHV) with pressures ranging from $5 \times 10^{-8}$ Torr to $3 \times 10^{-10}$ Torr. The total instrumental energy resolution was $\approx 0.2$ eV for the Fe edge, and $\approx 1.2$ eV for the As edge. Photon energies were calibrated either by simultaneously measuring standard compounds which intercepted a tiny fraction of the monochromatic beam, or with As $3d_{5/2}$ core level photoemission measurements.

The measurements were carried out at normal and near grazing incidence, and by tuning the source between horizontal (H) and vertical (V) linear photon polarization. At normal incidence, the electric field vector for H and V polarization is oriented in the sample plane (*ab* plane). At grazing incidence, the electric field vector is oriented along the surface normal (*c* axis) for H polarization, whereas it remains in the *ab* plane for V polarization. Based on Laue measurements, the samples were mounted such that the V polarization was maintained collinear with the Fe-Fe direction, i.e. the [110] direction in tetragonal notation. In what follows, we show Fe and As $L_{2,3}$ spectra collected in TEY on the $BaFe_2As_2$ parent compound below $T_{ST}$. Within the capabilities of our experimental set-up, Fe and As $L_{2,3}$ spectra collected in the optimally doped $BaFe_{1.8}Co_{0.2}As_2$ are effectively identical to those of the parent compound. Identical conclusions are reached when comparing spectra collected in TFY. The spectra have been normalized at the background on the low photon energy side. Other normalization methods do not change the overall results, thus corroborating the robustness of the data analysis.

**RESULTS AND DISCUSSION**

A direct comparison of Fe $L_{2,3}$ edge spectra collected in TEY at grazing incidence in the pristine and in-situ cleaved at $5 \times 10^{-8}$ Torr parent compound $BaFe_2As_2$ above $T_{ST}$ is shown in Fig. 1a for V and H polarizations, with the light polarization vector lying in-plane and out-of-plane, respectively. Spectral features which are particularly enhanced in V or H polarization correspond to states with in-plane or out-of-plane character, respectively.

The main difference between the pristine and the cleaved samples consists in the presence in the former of a structure at 710 eV, particularly pronounced for H polarization, indicative of surface contamination in the form of Fe oxide as a consequence of exposure to air. The polarization dependence of this feature, also present in the TFY measurements, is not due to artifacts resulting from different penetration depths since the V and H polarizations sample the same amount of material. Rather, it indicates that the Fe-O bonds originating from the surface contamination are of predominant out-of-plane character, a point to which we will further comment below. Although this oxide feature is markedly reduced for the cleaved samples, extreme care has to be exercised in assuring that the spectra obtained on freshly cleaved surfaces are representative of the intrinsic electronic structure of the compounds. Strikingly, the surfaces are found to be very reactive, even when freshly exposed in UHV at pressures in the $10^{-9}$ Torr range, with signatures of oxygen contamination, notwithstanding the absence of obvious signatures of it in the Fe $L_{2,3}$ edge spectra such as those in Fig. 1a. As an example, Fig. 1b shows XAS O K edge spectra taken on a BaFe$_{1.8}$Co$_{0.2}$As$_2$ sample freshly cleaved in UHV at $2 \times 10^{-9}$ Torr. For H polarization, an O K signal is detected at both grazing and normal incidence. These data reveal interactions of the freshly cleaved surface with molecules containing O in the vacuum system, most likely H$_2$O, consistent with the presence of Fe-O bonds as suggested by the presence of the Fe oxide peak visible in the pristine sample at 710 eV in the Fe L$_{2,3}$ spectra shown in Fig. 1a. Interestingly, we find that the As L$_3$ edge spectra do not exhibit changes in the lineshape when comparing pristine samples with thick native oxide surface layers and sample cleaved in UHV, further suggesting that the O detected on the surface is bonded to the Fe atoms. Although the low signal intensity does not allow the carrying out of a detailed polarization analysis of the O K edge spectra, the polarization dependence of the Fe L$_{2,3}$ spectra for the pristine sample (cf. Fig. 1a) suggests that the O–Fe bonds are primarily out-of-plane oriented. Cleaving samples *in-situ* at pressures better than $5 \times 10^{-10}$ Torr was essential in order to minimize the surface contamination and obtain spectra truly representative of the intrinsic electronic structure of these compounds, with no indication of Fe oxide peaks and absence of detectable O K edge signal, as all of the Fe spectra shown hereafter.

Fig. 2 shows a direct comparison of Fe L$_3$ edge spectra at different geometries for BaFe$_2$As$_2$ surfaces freshly cleaved in UHV ($3 \times 10^{-10}$ Torr). The main peak, centered at 708.5 eV, is broadened by the presence of spectral weight at $\approx$ 707.6 and $\approx$ 710 eV. A weak and $\approx$ 2 eV - wide structure is also visible at $\approx$ 712 eV. The spectral weight at $\approx$ 710 eV, which is typically present in compounds with Fe 3d-X np hybridization (X denotes an sp element) such as FeAs or FeSi, is indicative of the covalent nature of the Fe and As states, in analogy with the Fe XAS spectra in oxypnictide 1111 systems [15]. We stress that this spectral weight is intrinsic, and not due to spectral signatures related to Fe oxide, as confirmed by the absence of detectable O K edge signal.

Because of the absence of additional satellite peaks typically present when correlation effects are at work, the Fe spectral lineshape appears noticeably different compared to that of strongly correlated Fe oxides, and more akin to that of Fe metal, consistent with a delocalized character of the 3d states and indicative of the occurrence of weak electronic correlations [15,16]. The weak nature of the electronic correlations is also suggested by the agreement resulting from directly comparing the data with the XAS spectrum calculated with partial density of states (p-DOS) and modulated by matrix elements (Fig. 2) [16]. We stress that spectra collected at different temperatures above and below $T_{ST}$, or in the doped compound $BaFe_{1.8}Co_{0.2}As_2$ in which magnetism is absent, exhibit the same features, thus validating our theoretical approach, which does not include magnetism. This calculation places the main XAS peak at the correct position $\approx$ 0.9 eV above the Fermi level $E_F$ (h$\nu$ = 708.5 eV), in much better agreement than DMFT calculations with substantial Hubbard U, which place the peak much closer to $E_F$ [17,18]. The calculated spectrum exhibits spectral weight at h$\nu \approx$ 707.6 eV in proximity of $E_F$, and two peaks at 710.3 eV ($\approx$ 2.5 eV above $E_F$) and at $\approx$ 712 ($\approx$ 4 eV above $E_F$), which match the positions of the broad shoulders in the Fe $L_{2,3}$ spectra. As we discuss below, these structures originate from overlap of Fe s/d and As s/p states.

We now turn on exploring the symmetry and topology of the Fe d orbitals. Fe $L_3$ edge spectra in $BaFe_2As_2$ above $T_{ST}$ are shown in Fig. 2 at normal and near-grazing incidence for V and H polarizations. Single crystals formdomains with orthogonal orientations below $T_{ST}$, with the *a* and *b* axes rotated by 90° [19], i.e. some domains have the *a* axis aligned with vertical polarization and other domains have the *b* axis aligned with vertical polarization. Consistent with this observation, we detect no linear dichroism effects, that is, the spectra collected at normal incidence for H and V polarizations are identical. At grazing incidence, we observe a dichroic effect primarily visible as a difference in the amplitude of the main peak (708.5 eV) for V and H polarizations. In this geometry, as the polarization is rotated from V to H, the electric field vector is projected from the surface plane onto the surface normal, thus allowing the discrimination of in-plane and out-of-plane character of the orbitals. Spectral features which are particularly enhanced in V or H polarization correspond to states with in-plane or out-of-plane character, respectively.

In Fig. 2 we compare the difference of the V and H spectra, i.e. the dichroic signal, with the orbital projections of the Fe-s and Fe-d p-DOS, since these are the orbitals involved in the XAS process. The calculation accounts for the polarization dependence of the main peak at $\approx$ 1 eV above $E_F$. Both in plane $d_{x2-y2}$ and out of plane $d_{z2}$ Fe states are correctly placed in correspondence of the main peak, as confirmed by the data showing that when the electric field vector is rotated from in-plane to out-of-plane (V and H polarizations), the difference in XAS intensities reflects the change in spectral weight of the $d_{x2-y2}$ and $d_{z2}$ Fe orbitals. The calculation assigns the structure at $\approx$ 2.5 eV above $E_F$ to an orbital that is given by strong overlap between Fe $d_{xz+yz}$ and As $p_{x+y}$ states, further indicating that the shoulder at $\approx$ 710 eV in the Fe spectra is a fingerprint of the Fe - As hybridization. Most of the spectral weight originates from the Fe $d_{xz+yz}$ orbitals. If the $d_{xz}$ and $d_{yz}$ orbitals were isotropic and equally populated, one should expect an increase in spectral weight for out of plane polarization [20]. As indicated by the calculations, the spectral weight derived from $d_{xz}$ and $d_{yz}$ orbitals is very small and spread out in energy, thus rendering the detection of this effect particularly challenging. Indeed, the signal in proximity of the structure at 710 eV is very low, comparable to the noise level, thus making impossible to draw any conclusion on the true nature of the dichroism of the structure at 710 eV within the limits of the experimental resolution of our experiment. Two weaker dichroic signals with in-plane character are detected in proximity of $E_F$ (707.6 eV) and at $\approx$ 4 eV above $E_F$ (712 eV). The signal at $E_F$ appears to originate mainly from $d_{xz+yz}$ and $d_{xy}$ states, but only the latter are the likely cause of the weak in-plane dichroism. The peak at 712 eV originates mainly from Fe s states, with the dichroism being likely produced due to overlap with in-plane As $p_{x+y}$ states centered at $\approx 5$ eV.

One of the most striking observations in this study is the nature of the unoccupied As states, shown in Fig. 3. It is remarkable to observe such defined spectral features extending more than 25 eV above $E_F$. The As $L_3$ edge spectrum consists of five main structures particularly evident in the spectra collected with H polarization. The calculated XAS spectrum (p-DOS modulated by matrix element) correctly predicts the energy position of these structures.

The polarization dependence of the XAS spectra allows the determination of the As orbital topology. Similarly to Fe, spectra with H and V polarization at normal incidence are identical, as expected from the orthogonal nature of the domains [19]. Spectra collected at grazing incidence exhibit a marked dependence on the polarization, as particularly emphasized by the dichroic spectrum corresponding to the difference between V and H polarizations. The features of the dichroic spectrum can be explained by a direct inspection of the orbital projections of the As s/d p-DOS. Most of the spectral weight originates from the As $d_{xz+yz}$ orbital, but the dichroism emanates from the marked difference between out-of-plane As $d_{z2}$ orbital and in-plane As $d_{xy}$ and $d_{x2-y2}$ states. In particular, the structures located at $\approx 6$ eV, $\approx 15.6$ eV and $\approx 22$ eV above $E_F$ correspond to $d_{z2}$ states, while those at $\approx 9$ eV and $\approx 17$ eV are assigned to both $d_{xy}$ and $d_{x2-y2}$ states. Notably, the structure at $\approx 4$ eV above $E_F$, which appears to mainly originate from As s states, is located in correspondence to the broad shoulder in the Fe $L_3$ spectra and the As $p_{x+y}$ states (cf. Fig. 2), possibly indicating that As s and $p_{x+y}$ orbitals are hybridized, and overlap with Fe s states. The possibility of ascribing the spectral features of the data to the orbital projections of the p-DOS provides information about the bonding topology and supports simple qualitative geometrical considerations. The $d_{z2}$ orbitals are dumbells oriented along the z axis, while the $p_{x+y}$ orbital has lobes lying the in ab plane. Tetrahedral bonding is associated with the positive and negative lobes of the Fe $d_{xz+yz}$ orbital overlapping the As $p_{x+y}$ orbital to give bonding and antibonding states, respectively. XAS probes the antibonding orbitals with higher orbital energy than the bonding states, which are located several eV deep in the valence band. Interestingly, it would appear that the unoccupied states do not exhibit the renormalization effects, which affect the occupied electronic bands below $E_F$ as exposed by several ARPES investigations. However, it should be emphasized that we are probing a much wider energy range, extending beyond the energy of the main d bands, which may well be renormalized.

It thus appears that there is considerable agreement between XAS data and DFT predictions for both Fe and As edges. In particular, the XAS spectra at both Fe and As edges are well matched by the unoccupied DOS, indicating the occurrence of weak to moderate electronic correlations, as also suggested by the Fe spectral lineshape, which is similar to that of Fe in compounds with a delocalized character of the 3d states. The data and calculations indicate a substantial degree of Fe-As hybridization, as revealed by the strong overlap of Fe $d_{xz+yz}$ states with As $p_{x+y}$ states. This is particularly important since the Fe $d_{xz+yz}$ states contribute substantially to the DOS at $E_F$.

In summary, the elemental sensitivity of polarization dependent XAS has been exploited for determining the orbital symmetry of electron doped iron-arsenide superconductors $Ba(Fe_{1-x}Co_x)_2As_2$. Our analysis reveals a substantial degree of Fe-As hybridization related to the bonding topology of the Fe $d_{xz+yz}$ states, which contribute substantially to the DOS at the Fermi level. We also find results in accord with DFT predictions of the orbital characters away from $E_F$. This means that despite the known problems of DFT calculations, such as a strong overestimation of the magnetic tendency of these materials and difficulties describing the interplay between magnetism and Fe-As bonding [21], the orbital occupancies, their relative energies in spectra, and the interplay with ligand orbitals are not strongly modified by correlations. These findings are quite different from what is expected in correlated oxides, such as cuprates, and impose stringent constraints on theories capable of providing a correct description of FeSC materials.

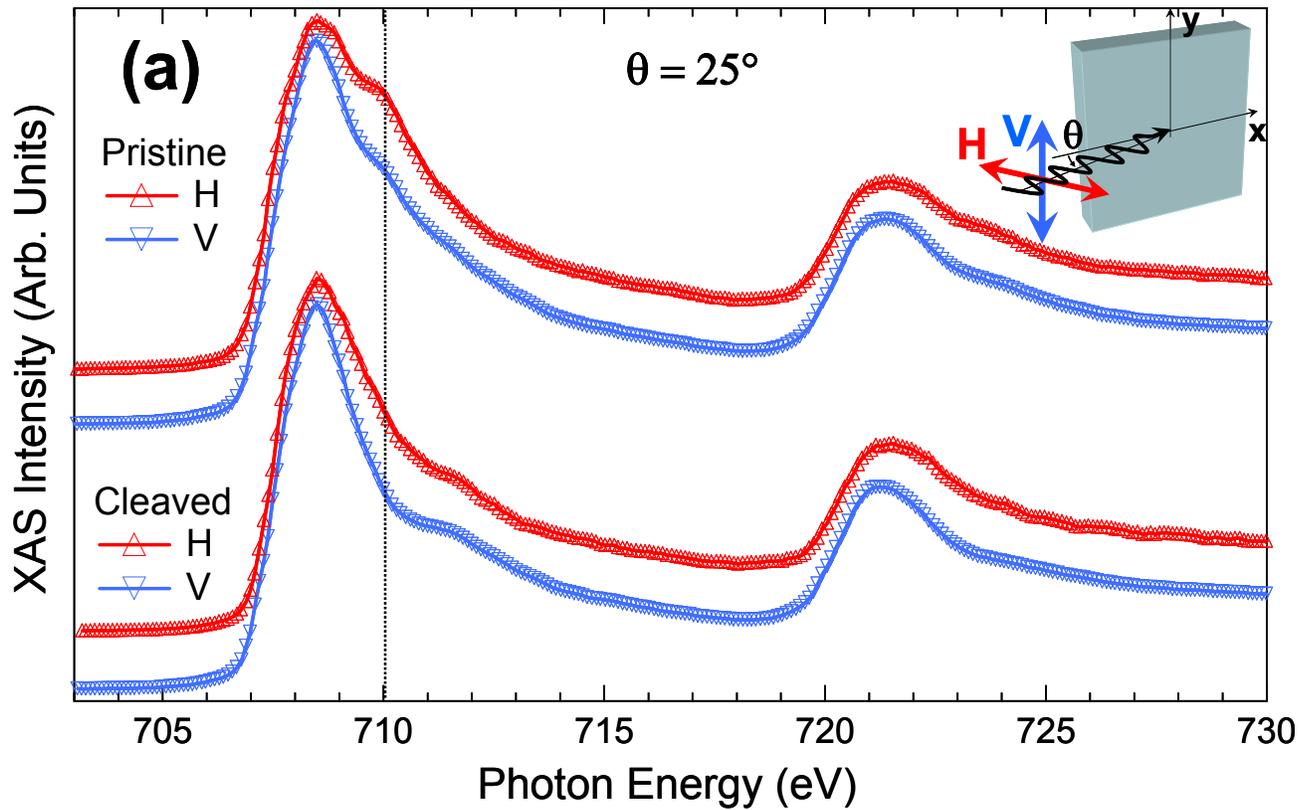
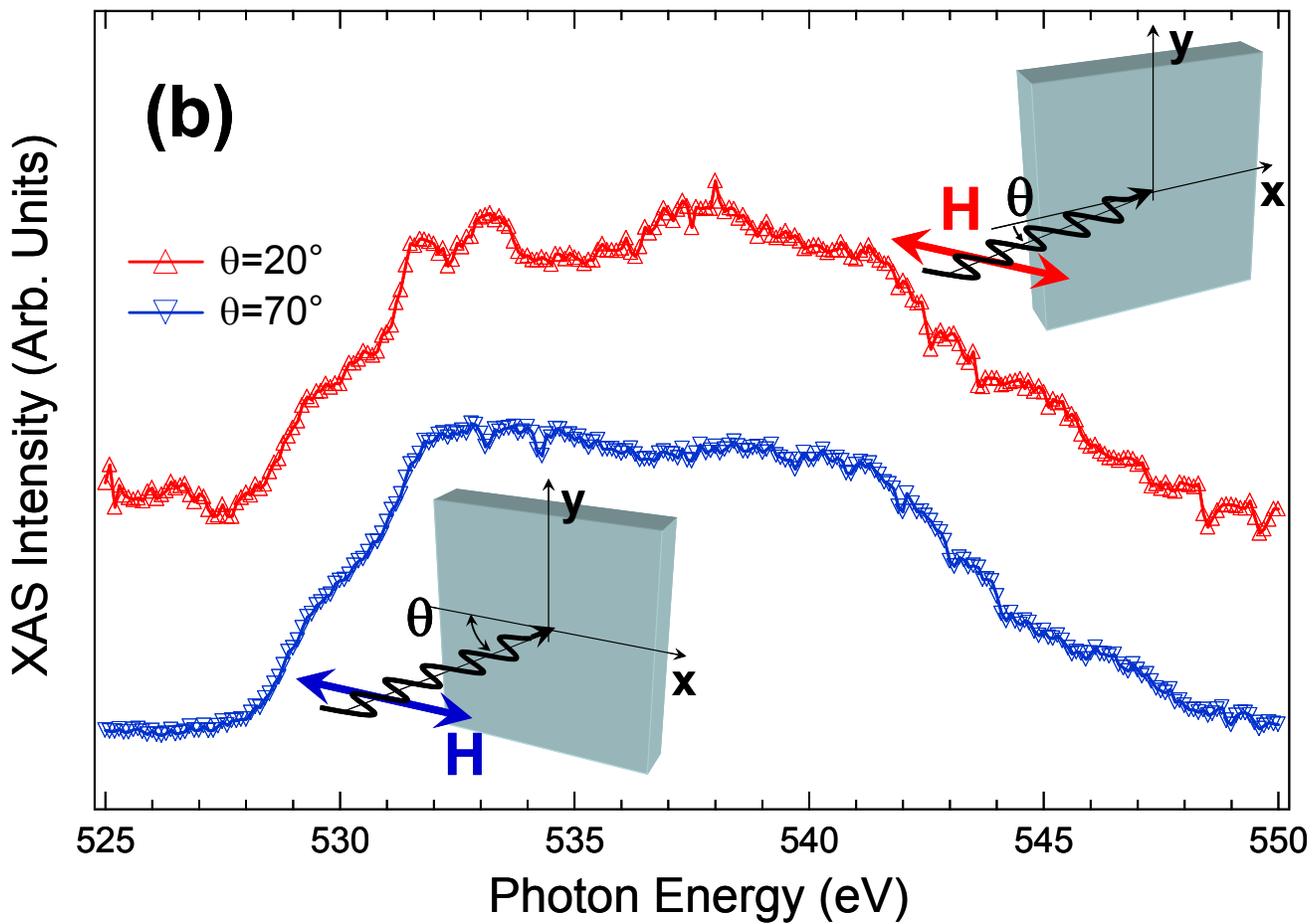

Fig. 1 (color online)   Fe $L_{2,3}$ edge and O K edge XAS spectra illustrating the effects of surface contamination. (a) Fe $L_{2,3}$ edge XAS spectra measured in TEY at 200 K for a pristine oxide surface and a freshly cleaved surface in UHV ($5 \times 10^{-8}$ Torr) of the $BaFe_2As_2$ parent compound. The incident beam is at grazing incidence ($\theta = 25°$) to the surface with the electric field polarized in the horizontal (H) and vertical (V) directions, as denoted by the double headed arrows in the inset. The x and y axes denote the direction of the Fe-Fe bonds. The presence of an additional Fe oxide peak visible in the pristine sample at 710 eV is indicative of surface contamination. (b) XAS O K edge spectra taken on a $BaFe_{1.8}Co_{0.2}As_2$ sample freshly cleaved in UHV ($2 \times 10^{-9}$ Torr) at room temperature, but still showing occurrence of contamination. The presence of the O K edge signal indicates an immediate interaction of the freshly cleaved surface in UHV with molecules containing O in the vacuum system (see text), a token of the high reactivity of the surfaces of these compounds, even when freshly exposed in UHV at pressures in the low $10^{-9}$ Torr range.

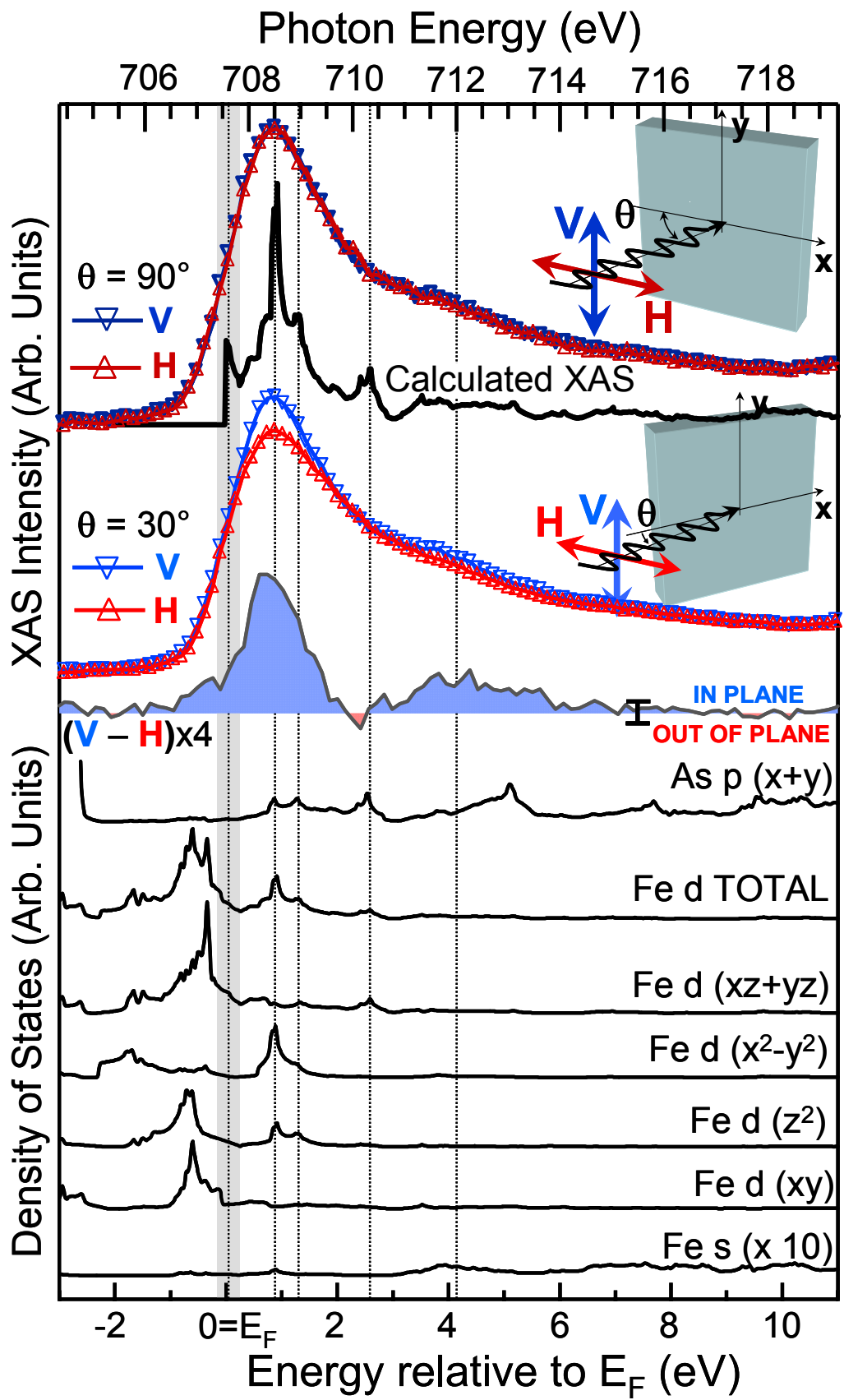

Fig. 2 (color online)   Fe $L_3$ edge XAS spectra measured in TEY for $BaFe_2As_2$ surfaces freshly cleaved in UHV ($3 \times 10^{-10}$ Torr) and comparison with theory. The insets show the schematic layout of the experimental geometries. The incident beam is at normal ($\theta = 90°$) and grazing incidence ($\theta = 25°$) to the surface with horizontal (H) and vertical (V) polarizations, as denoted by the double headed arrows.  The x and y axes denote the direction of the Fe-Fe bonds.  The thick black line denotes the XAS spectrum calculated using orientation averaged matrix elements and DOS calculated with DFT without magnetism.  The curve denoted as "V-H" is the dichroic signal obtained by subtracting the H polarization spectra from the V polarization spectra at grazing incidence.  The error bar of the amplitude of the dichroic signal, roughly equal to twice the noise level in the XAS spectra, is indicated on the right of the "V-H" curve.  Bottom: orbital projections of the Fe-s and Fe-d p-DOS calculated with DFT.  The As $p_{x+y}$ states are also indicated to illustrate the hybridization with the Fe s and Fe $d_{xz+yz}$ states.

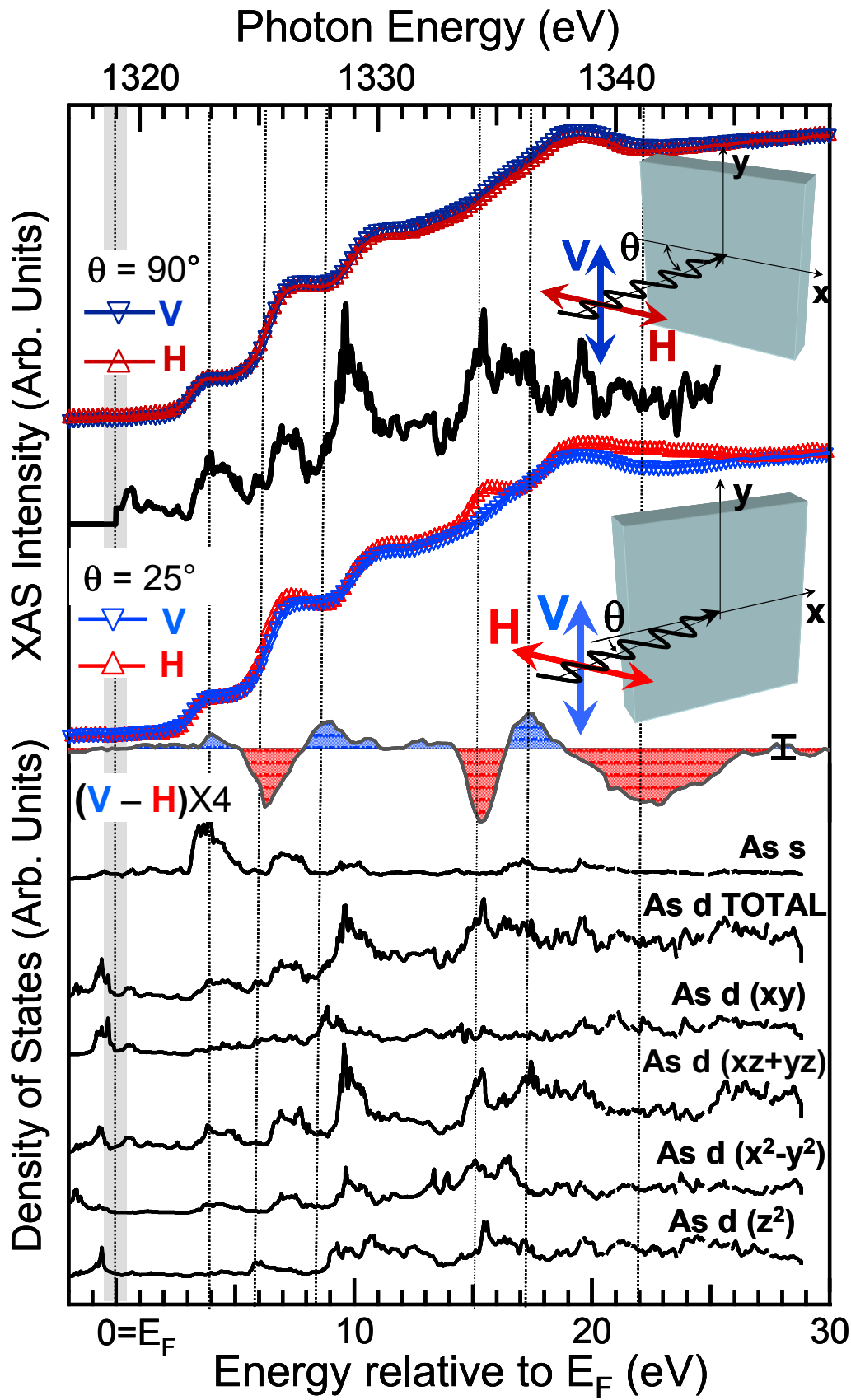

Fig. 3 (color online) As $L_3$ edge XAS spectra measured in TEY for $BaFe_2As_2$ surfaces freshly cleaved in UHV ($5 \times 10^{-8}$ Torr) and comparison with theory. The insets denote the experimental geometries, as in Fig. 2. The thick black line denotes the XAS spectrum calculated using orientation averaged matrix elements and DOS calculated with DFT without magnetism. The curve denoted as "V-H" is the dichroic signal. Bottom: orbital projections of the As-s and As-d p-DOS calculated with DFT. The error bar of the amplitude of the dichroic signal, roughly equal to twice the noise level in the XAS spectra, is indicated on the right of the "V-H" curve.

## ACKNOWLEDGMENTS

The work at the ALS, APS and Elettra is supported by NSF grant DMR-0804902. The work at Oak Ridge is sponsored by the Department Of Energy, Division of Materials Sciences and Engineering. Oak Ridge National Laboratory is managed by UT-Battelle, LLC, for the U.S. Department of Energy under Contract No. DE-AC05-00OR22725. Work at Argonne is supported by the U.S. Department of Energy, Office of Science, under Contract No. DE-AC02-06CH11357. A. S. S and M. A. McG. acknowledge support from the Eugene P. Wigner Fellowship at ORNL.